\documentclass[aps,preprint,showpacs]{revtex4}
\usepackage{epsfig}

\begin{document}

\title{Decoherence of charge states in double quantum dots due to cotunneling}

\author{U.\ Hartmann}
\email{hartmann@theorie.physik.uni-muenchen.de}
\author{F.K.\ Wilhelm}

\affiliation{Sektion Physik and CeNS, Ludwig-Maximilians-Universit\"at, 
Theresienstr.\ 37, 80333 M\"unchen, Germany}

\begin{abstract}
Solid state quantum bits are a promising candidate for the realization of a
{\em scalable} quantum computer, however, they are usually strongly limited
by decoherence.
We consider a double quantum dot charge qubit, whose
basis states are defined by the position of
an additional electron in the system of two laterally coupled quantum dots.
The coupling of these two states can  be controlled externally by a
quantum point contact between the two dots. We discuss the decoherence
through coupling to the electronic leads due to cotunneling processes.
We focus on a simple Gedanken experiment, where the system is initially
brought into a superposition and then the inter-dot coupling is removed
nonadiabatically. We treat the system by invoking the Schrieffer-Wolff 
transformation in order to obtain a transformed Hamiltonian describing 
the cotunneling, and then obtain the dynamics of the density matrix 
using the Bloch-Redfield theory. As a main result, we show that 
there is energy relaxation even in the absence
of inter-dot coupling. This is in contrast to what would be expected from
the Spin-Boson model and is due to the fact that a quantum dot
is coupled to {\em two} distinct baths. 
\end{abstract}


\maketitle

Quantum dots (``artificial atoms'') are prototype systems for studying
the properties of discrete levels embedded in a solid-state environment
\cite{Ashoori}. 
In particular, various schemes for realizing quantum bits, 
fully controlled quantum coherent two-state systems, using quantum dots have
been brought forward. Next to using optically excited charge states in  
quantum dots \cite{Rossi} and electronic quantum dots used
for spin manipulation \cite{Loss}, it has been proposed \cite{Blick}
to use the charge states of a double quantum dot as a computational basis.
%Einleitung ausgebaut zum ``setting into context'' 
\begin{figure}
\begin{center}
\epsfxsize=8cm
\epsfbox{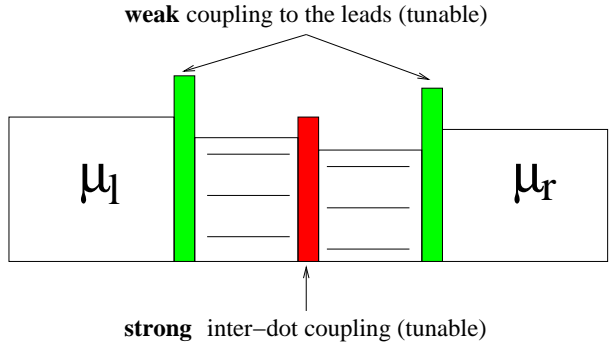}
\end{center}
\caption{Sketch of the double dot system. The coupling of the double 
dot to the leads is assumed to be weak, whereas the coupling between the dots
can be strong. The leads are biased such that sequential tunneling is 
supressed.}
\label{dots}
\end{figure}
The proposed setup is sketched in Figure~\ref{dots}.
In order to minimize the inevitable decoherence through coupling
to the electronic leads, the system can be brought into the
Coulomb blockade regime where sequential tunneling is supressed. We are 
going to discuss in this article, how
the inevitable cotunneling still decoheres the system in this regime. 
The calculation is carried out
for one specific Gedanken experiment which should capture the most 
generic features, the decay of a superposition state when
the coupling between the dots is switched off. A more complete treatment
of this setup is in preparation \cite{UdoFrankPrep}. 

 We
restrict our analysis on spin-polarized electrons.
The relevant Hilbert space is characterized by four basis states, written 
as $\vert i,j\rangle$, which denotes $i$ additional
electrons on the left dot, $j$ additional electrons on the right dot.
The two states $\vert 1,0\rangle$ and
$\vert 0,1\rangle$ define the computational basis \cite{Nielsen}. 
In order to describe cotunneling, we use the closest energetically
forbidden
states as virtual intermediate states. These are
 $\vert v_0\rangle=\vert 0,0\rangle$ and $\vert v_2\rangle=\vert 1,1\rangle$. 
Zero- and two
electron states with internal polarization are energetically even
less favorable 
due to the high charging energy of the individual dots. 

The Hamiltonian of this system can be written as
\begin{eqnarray}
H & = & H_{0} + H_{1}\\
H_{0} & = & \epsilon_{\rm as} (\hat{n}_l-\hat{n}_r) -
\epsilon_\alpha \hat{n}_{v_0}+\epsilon_\beta \hat{n}_{v_2}\nonumber \\
& &{} + \gamma \sum_{n} (a^{L\dagger}_{n}a^{R}_{n}+a^{R\dagger}_{n}a^{L}_{n})
+ \sum_{\vec{k}} \epsilon^{L}_{\vec{k}}b^{L\dagger}_{\vec{k}}b^{L}_{\vec{k}} + \sum_{\vec{k\prime}} \epsilon^{R}_{\vec{k\prime}}b^{R\dagger}_{\vec{k\prime}}b^{R}_{\vec{k\prime}}\\
H_{1} & = & t_{c} \sum_{\vec{k},n} (a^{L\dagger}_{n}b^{L}_{\vec{k}}+a^{L}_{n}b^{L\dagger}_{\vec{k}}) + t_{c} \sum_{\vec{k\prime},m} (a^{R\dagger}_{m}b^{R}_{\vec{k\prime}}+a^{R}_{m}b^{R\dagger}_{\vec{k\prime}})
\end{eqnarray}
Note, that the sum over dot states only runs over the restricted Hilbert
space described above.
$H_{0}$ describes the energy spectrum of the uncoupled system,
whereas the tunneling part $H_{1}$ describes the
coupling of each dot to its lead and will be treated as a perturbation.
$\hat{n}_{l/r}$ are the number operators for the additional
electrons on either dot. The asymmetry energy $\epsilon_{\rm as}$ describes
the difference between the energy level for the additional electron in left dot
and the corresponding energy level in the right dot. It can be tuned
through via the gate voltages which are applied at each dot.
$\epsilon_{\beta}$ and $\epsilon_{\alpha}$ are the energy differences
towards the higher level $\vert v_2\rangle$ and 
the lower level $\vert v_0\rangle$ respectively.
$\gamma$ is the tunable inter-dot coupling.
The $a^{(\dagger)}$s and $b^{(\dagger)}$s denote the creation/destruction 
operators
in the dots and leads.
In $H_{1}$ the symbol $t_{c}$ represents the coupling constant concerning the
coupling of the dots to the leads, which should be small compared to the
asymmetry energy. Note, that we have chosen a
slightly asymmetric notation in order to highlight the physical model: For
the actual calculation, $H_1$ is also expressed in the eigenstate basis
of the dot.

For our Gedanken experiment, we assume that first the inter-dot coupling 
$\gamma$ is large ($\gamma\gg \epsilon_{\rm as},V$) such that the system 
relaxes into the ground state, which
is a molecular superposition state of the form 
$|g\rangle=(|0,1\rangle-|1,0\rangle)/\sqrt{2}$. Then the gate voltage that 
controls
the inter-dot coupling is switched to high values,
so that the coupling is practically
zero. After this, the system dephases and relaxes into a thermal mixture of 
the localized
eigenstates of the new system.

Thus, in order to describe decoherence, 
we only have to consider the case $\gamma=0$ K. This means, that
 $H_{0}$ is already diagonal, i.e.\
the states $\vert 1,1\rangle$, $\vert 1,0\rangle$, $\vert 0,1\rangle$ and $\vert 0,0\rangle$ are 
eigenstates of our system.

The decoherence is analyzed applying the well-established 
Bloch-Redfield 
theory,
which is based on the Born approximation in the system-bath coupling. As we 
are 
in the
Coulomb blockade regime, the rates evaluated from the 
original coupling Hamiltonian 
$H_1$ vanish in that order. In order to treat cotunneling
with this formalism, we perform a generalized Schrieffer-Wolf transformation
\cite{swtrafo,Cohen}. This transformation maps our 
original Hamiltonian $H_1$, which is zero in the computational basis
but couples the computational states to the $|v_{0/2}\rangle$ onto 
a Hamiltonian which does {\em not} 
have this coupling to higher states but which 
has nonzero matrix elements in the computational basis. 
The new terms in the Hamiltonian
describe the amplitude of transitions between the basis states via the
intermediate states. We perform this transformation perturbatively 
up to second order,
i.e.\ all processes involving at most {\em one} intermediate state are
taken into account. 

The new Hamiltomian $H_{I}$ in our special case then can 
be written as
\begin{eqnarray}
H_{I,++} & = & A(R\dagger,R,++)\ b^{R\dagger}_{m}b^{R}_{n}+A(L,L\dagger,++)\ b^{L}_{l}b^{L\dagger}_{k}\\
H_{I,--} & = & A(L\dagger,L,--)\ b^{L\dagger}_{k}b^{L}_{l}+A(R,R\dagger,--)\ b^{R}_{n}b^{R\dagger}_{m}\\
H_{I,+-} & = & A(R\dagger,L,+-)\ b^{R\dagger}_{m}b^{L}_{l}+A(L,R\dagger,+-)\ b^{L}_{l}b^{R\dagger}_{m}\\
H_{I,-+} & = & A(L\dagger,R,-+)\ b^{L\dagger}_{k}b^{R}_{n}+A(R,L\dagger,-+)\ b^{R}_{n}b^{L\dagger}_{k}.
\end{eqnarray}
The $+$ and $-$ signs are indizes for the states $\vert 1,0\rangle$ resp.
$\vert 0,1\rangle$. We call the $A$s Schrieffer-Wolff coefficients, they
are calculated along the lines of \cite{Cohen} using mainly second order
perturbation theory.
For example, $A(R\dagger,R,++)$ is
\begin{equation}
A(R\dagger,R,++)=\frac{t^{2}_{c}}{2}\Bigg \lbrack \frac{1}{\epsilon^{R}_{m}-(-\epsilon_{\rm as} + \epsilon_{\beta})} - \frac{1}{\epsilon^{R}_{n} -(\epsilon_{\rm as} - \epsilon_{\beta})}\Bigg \rbrack .
\end{equation}
We now use the Bloch-Redfield equations \cite{Weiss,Blum}
\begin{equation}
\dot{\rho}_{nm}(t) = -i\omega_{nm}\rho_{nm}-\sum_{k,l} R_{nmkl}\rho_{kl}(t)
\label{redfield}
\end{equation}
where $R_{nmkl}$ are the elements of the Redfield tensor.
These equations of motion for the
reduced density matrix are obtained within Born approximation in the 
effective system-bath coupling, so after the Schrieffer-Wolff transformation,
$R$ is of order $t_c^4$. Let us remark that our perturbation theory 
naturally breaks down below the Kondo
temperature, which can however be made arbitrarily small by lowering $t_c$ 
through pinching
off the contacts to the reservoirs. 

The Bloch-Redfield equations are of Markovian form, however, by properly
using the free time evolution of the system, they take into account all bath
correlations which are relevant within the Born approximation
\cite{Hartmann}.

The Redfield tensor has the form
\begin{equation}
R_{nmkl}=\delta_{lm} \sum_{r} \Gamma^{(+)}_{nrrk} + \delta_{nk} \sum_{r} \Gamma^{(-)}_{lrrm} - \Gamma^{(+)}_{lmnk} - \Gamma^{(-)}_{lmnk}.
\end{equation}
The rates entering the Redfield tensor elements are given
by the following Golden-Rule expressions
\begin{eqnarray}
\Gamma^{(+)}_{lmnk} & = & \hbar^{-2} \int_{0}^{\infty} \limits dt \ e^{-i\omega_{nk}t}
\langle \tilde{H}_{I,lm}(t)\tilde{H}_{I,nk}(0) \rangle 
\label{eq:plusrate}\\
\Gamma^{(-)}_{lmnk} & = & \hbar^{-2} \int_{0}^{\infty} \limits dt \ e^{-i\omega_{lm}t}
\langle \tilde{H}_{I,lm}(0)\tilde{H}_{I,nk}(t) \rangle
\label{eq:minusrate}
\end{eqnarray}
where $H_I$ appears in the interaction representation (written as $\tilde{H}_{I}$).
In our formalism, it is of crucial importance that the expectation
values over $H_I$ vanish, i.e. that the bath produces only noise.
As a number of expectation values of $H_{I}$ turns out to be finite, we tacitly
replace $H_I$ by $H_I-\langle H_I\rangle$ in eqs.\ (\ref{eq:plusrate})
and (\ref{eq:minusrate}) and use the 
finite expectation values to renormalize the
diagonalized, unperturbed Hamiltonian 
${H}_{0}\rightarrow H_0+\langle H_I\rangle$. In our case, 
the effect of this renormalization is of the order of $0.1$ \% of the original matrix elements of $H_{0}$.\\
After a straightforward calculation of the above Golden-Rule rates, one gets
in the general case a (large) sum over terms with the generic form
\begin{eqnarray}
\Gamma^{(+)} & = & c \ \Bigg \lbrace \frac{i\pi}{\epsilon_{b}-\epsilon_{a}\mp 2\epsilon_{\rm as}}\lbrack f_{1}(\epsilon_{b}\mp 2\epsilon_{\rm as})(1-f_{2}(\epsilon_{b}))-f_{1}(\epsilon_{a})(1-f_{2}(\epsilon_{a}\pm 2\epsilon_{\rm as}))\rbrack + {} \nonumber \\
& & {} + \frac{-n_{1}(\mu_{2}\mp 2\epsilon_{\rm as})}{\epsilon_{b}-\epsilon_{a}\mp 2\epsilon_{\rm as}} \Bigg \lbrack \psi \Bigg (\frac{1}{2}+\frac{i\beta}{2\pi}\Bigg (\epsilon_{b}\mp 2\epsilon_{\rm as} -\mu_{1}\Bigg )\Bigg ) - \psi \Bigg (\frac{1}{2}+\frac{i\beta}{2\pi}\Bigg (\epsilon_{a}-\mu_{1} \Bigg )\Bigg ) - {} \nonumber \\
& & {} - \psi \Bigg (\frac{1}{2}+\frac{i\beta}{2\pi}\Bigg (\epsilon_{b}-\mu_{2}\Bigg )\Bigg ) + \psi \Bigg (\frac{1}{2}+\frac{i\beta}{2\pi}\Bigg (\epsilon_{a}\pm 2\epsilon_{\rm as} -\mu_{2}\Bigg )\Bigg )  \Bigg \rbrack \Bigg \rbrace\\
\Gamma^{(-)} & = & c \ \Bigg \lbrace \frac{i\pi}{\epsilon_{b}-\epsilon_{a}\mp 2\epsilon_{\rm as}}\lbrack f_{2}(\epsilon_{b})(1-f_{1}(\epsilon_{b}\mp 2\epsilon_{\rm as}))-f_{2}(\epsilon_{a}\pm 2\epsilon_{\rm as})(1-f_{1}(\epsilon_{a}))\rbrack + {} \nonumber \\
& & {} + \frac{-n_{2}(\mu_{1}\pm 2\epsilon_{\rm as})}{\epsilon_{b}-\epsilon_{a}\mp 2\epsilon_{\rm as}} \Bigg \lbrack -\psi \Bigg (\frac{1}{2}+\frac{i\beta}{2\pi}\Bigg (\epsilon_{b}\mp 2\epsilon_{\rm as} -\mu_{1}\Bigg )\Bigg ) + \psi \Bigg (\frac{1}{2}+\frac{i\beta}{2\pi}\Bigg (\epsilon_{a}-\mu_{1}\Bigg )\Bigg ) + {} \nonumber \\
& & {} + \psi \Bigg (\frac{1}{2}+\frac{i\beta}{2\pi}\Bigg (\epsilon_{b}-\mu_{2}\Bigg )\Bigg ) - \psi \Bigg (\frac{1}{2}+\frac{i\beta}{2\pi}\Bigg (\epsilon_{a}\pm 2\epsilon_{\rm as} -\mu_{2}\Bigg )\Bigg )  \Bigg \rbrack \Bigg \rbrace
\end{eqnarray}
where $c = \frac{t_{c}^{4}\pi V^{2}m_{\ast}^{2}}{4\hbar (2\pi \hbar^{2})^{2}}$. One can express the coupling to the leads $t_{c}$ by $t_{c}=\sqrt{\frac{g}{8\pi^{2}}}\cdot \frac{E_{F}}{n}$, where $g$ is a conductance in terms of the
quantum conductance,
 $E_{F}$ is the Fermi energy of the leads and $n$ is the number of
electrons in the leads. Consequently, $c$ is then changed to
$c=\frac{t_{c}^2g}{32 \pi \hbar}$. The $\epsilon_a$ and $\epsilon_b$ are terms
containing varying sums or differences of $\epsilon_{\beta}$, $\epsilon_{\alpha}$ and $\epsilon_{\rm as}$. Due to the multitude of possibilities
for virtual transitions, each element of the Redfield tensor 
 contains a number of terms of this
generic structure. 

In the above equations, the terms containing the Fermi function $f(\epsilon)$
only play a
role close to resonance and can be neglected in the Coulomb blockade
\cite{Koenig}.
The $n_{l/r}$'s represent Bose functions for the electron-hole pairs
(excitons) that are generated during the virtual processes.
The $\psi$'s denote Digamma functions and hence diverge logarithmically at
 low temperatures.

By solving equation (\ref{redfield}), one finds that 
the off-diagonal elements decay towards zero 
on a time scale $\tau_{\phi}$ (dephasing time) whereas the 
diagonal density matrix elements equilibrate on a
time scale $\tau_{r}$ (relaxation time).

\begin{figure}
\begin{center}
\epsfxsize=9cm
\epsfbox{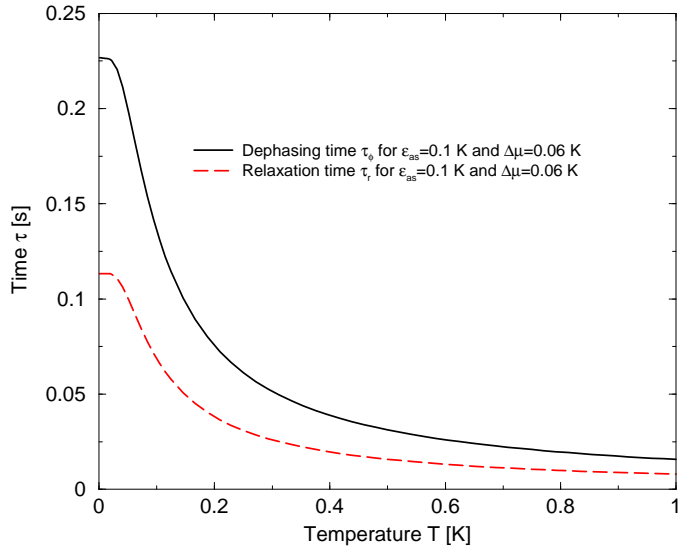}
\end{center}
\caption{Relaxation and dephasing time ($\tau_{r}$ and $\tau_{\phi}$) as
a function of temperature $T$, with $\mu_{l}=0.85$~K, $\mu_{r}=0.91$~K,
$\epsilon_{\rm as}=0.1$~K, $\epsilon_{\beta}=11$~K, $\epsilon_{\alpha}=9$~K,
$g=0.1$, $V=10^{-12} \ \rm{m^{2}}$, $E_{F}=5$~meV and
$n/V=1.7\cdot 10^{15}\ \rm{m^{-2}}$.}
\label{pss_T}
\end{figure}

Using the above expressions, we find the rates as
\begin{eqnarray}
\Gamma_{r} & = & 2 \ (\Gamma^{(+)}_{+--+}+\Gamma^{(+)}_{-++-})\\
\Gamma_{\phi} & = & \frac{\Gamma_{r}}{2} + (\Gamma^{(+)}_{++++}+\Gamma^{(+)}_{----}-2 \Gamma^{(+)}_{++--})
\end{eqnarray}
where
\begin{eqnarray}
\Gamma^{(+)}_{+--+} & = & \Gamma^{(-)}_{+--+} = c\ (-n_{r}(\mu_{l}+ 2 \epsilon_{\rm as}))Z \label{eqn1}\\
\Gamma^{(+)}_{-++-} & = & \Gamma^{(-)}_{-++-} = c\ (-n_{l}(\mu_{r}- 2 \epsilon_{\rm as}))Z \\
\Gamma^{(+)}_{++++} & = & \Gamma^{(-)}_{++++} = c\ \frac{1}{\beta} \ Y_{1} \\
\Gamma^{(+)}_{----} & = & \Gamma^{(-)}_{----} = c\ \frac{1}{\beta} \ Y_{-1} \\
\Gamma^{(+)}_{++--} & = & \Gamma^{(-)}_{++--} = c\ \frac{1}{\beta} \ Y_{1,-1} \rm . \label{eqn2}
\end{eqnarray}
$Z$ is a function containing several $\psi$-functions (or logarithms).
$Y_{1}$, $Y_{-1}$ and
$Y_{1,-1}$ are different functions of several $\psi$'-(Trigamma-) functions
(or reciprocals), however, these functions only have a very weak 
temperature dependence. 
The most important part of the temperature dependence comes in through
$\frac{1}{\beta}$ and in $n_{l/r}$ and is 
summarized in Figure~\ref{pss_T}.
We find in Figure~\ref{pss_T} that the temperature dependence is 
similar to the Spin-Boson case \cite{Grifoni}. This can be confirmed by inspection of 
the formulas
(\ref{eqn1})-(\ref{eqn2}):
For the relaxation rate, one has only Bose functions
taken at the finite amount energy which is dissipated.
In case of the dephasing rate, there are also
terms that are proportional to $T$, which represent
dephasing processes which do not change the
energy of the qubit, i.e.\ cotunneling processes which originate and
end in the same state. This explains the observed behaviour. Note, that
in the Spin-Boson case, where there is only one lead, the situation 
corresponding to our Gedanken-experiment (no tunneling between
the classical states) would correspond to pure dephasing, whereas
in our system
 relaxation is always possible by extracting an electron on one side
and adding one on the other side from the other lead.

The numerical values for the relaxation and
dephasing times are comparedly huge, on the order of 100 milliseconds
as compared to the experimentally measured times, which are 
in the order
of nanoseconds.
Other possibilities to explain the small decoherence time are phononic and/or
photonic baths \cite{Brandes1,Brandes2,Alex}, or the influence of
the whole electronic circuitry.

 We analyzed relaxation and dephasing processes in a system of two laterally
coupled quantum dots which is coupled
to two electronic (i.e. fermionic) baths. We showed that even in the
case of vanishing inter-dot coupling, the system's energy can relax,
unlike in the Spin-Boson model. On top of that, the temperature
dependence of the rates resembles that of the Spin-Boson model. We
identify, that this originates in the fact that the cotunneling rates are
mostly sensitive to the distribution function of excitons.

As a next step, the case where the inter-dot coupling $\gamma$
has finite values will be considered \cite{UdoFrankPrep}. 
We thank J.\ von Delft, L.\ Borda, J. K\"onig, R.H.\ Blick, A.W.\ Holleitner,
A.K.\ H\"uttel and E.M. H\"ohberger for clarifying discussions. 
We acknowledge financial support from ARO, Contract-Nr. P-43385-PH-QC.

\end{document}